\documentclass[aps,pre,floats,twocolumn,showpacs,superscriptaddress]{revtex4}

\usepackage{graphicx}
\usepackage{graphics}
\usepackage{amsmath}
\usepackage{amsfonts}
\usepackage{amssymb}
\usepackage{epstopdf}
\usepackage{makeidx}
\usepackage{epsfig}
\usepackage{bm}
\usepackage{times}
\usepackage[colorlinks=true,citecolor=blue,linkcolor=blue,linktocpage=true,pagebackref=false]{hyperref}
\usepackage{color,soul} 
\usepackage{float}
\usepackage[english]{babel}

\newcommand{\be}{\begin{equation}}
\newcommand{\ee}{  \end{equation}}
\newcommand{\ba}{\begin{eqnarray}}
\newcommand{\ea}{  \end{eqnarray}}

\begin{document}

\title{Robustness of cultural communities in an open-ended Axelrod's model.}

\author{Alexis R. Hern\'andez}
\affiliation{Instituto de F\'{\i}sica,
Universidade Federal do Rio de Janeiro, 22452-970 Rio de Janeiro, Brazil}
\affiliation{Instituto de Biocomputaci\'on y F\'{\i}sica de Sistemas Complejos (BIFI), Universidad de Zaragoza, 50018 Zaragoza, Spain}
\author{Carlos Gracia-L\'azaro}
\affiliation{Instituto de Biocomputaci\'on y F\'{\i}sica de Sistemas Complejos (BIFI), Universidad de Zaragoza, 50018 Zaragoza, Spain}
\author{Edgardo Brigatti}
\affiliation{Instituto de F\'{\i}sica,
Universidade Federal do Rio de Janeiro, 22452-970 Rio de Janeiro, Brazil}
\author{Yamir Moreno}
\affiliation{Instituto de Biocomputaci\'on y F\'{\i}sica de Sistemas Complejos (BIFI), Universidad de Zaragoza, 50018 Zaragoza, Spain}
\affiliation{Departamento de F\'{\i}sica Te\'orica. University of Zaragoza, Zaragoza E-50009, Spain}
\affiliation{ISI Foundation, Turin, Italy} 

\date{\today}

\begin{abstract}
We consider an open-ended set of cultural features in the Axelrod's model of cultural dissemination. By replacing the features in which a high degree of consensus
is achieved by new ones, we address here an essential ingredient of societies: 
the evolution of topics as a result of social dynamics and debate. Our results show that, once cultural clusters
have been formed, the introduction of new topics into the social debate has little effect on them, but
it does have a significant influence on the cultural overlap. Along with the Monte-Carlo simulations, 
we derive and numerically solve an equation for the stationary cultural overlap based on a
mean-field approach. Although the mean-field analysis reproduces qualitatively the characteristic
phase transition of the Axelrod's model, it underestimates the cultural overlap, highlighting
the role of the local interactions in the Axelrod's dynamics, as well as the correlations
between the different cultural features.

\end{abstract}

\maketitle

\section{Introduction}

Agent-Based Modeling has become one of the major techniques to study complex adaptive systems, being
currently a paradigm in fields as diverse as ecology, sociology, economics or engineering. The use of
agent-based models (ABM) \cite{macy2002factors,tesfatsion2006handbook} in the study of
social phenomena provides a powerful theoretical framework
that gives useful insights about the fundamental mechanisms at work in social systems. In ABMs, agents represent interacting
entities (for example, individuals or groups of individuals) and are
characterized by a set of internal states. In particular, in opinion ABMs, 
agents are provided with a set of opinion variables \cite{castellano2009statistical}. In  1977, Axelrod 
\cite{axelrod1997dissemination} proposed an ABM for the dissemination of culture based on the idea of homophily, \textit{i.e.}, the tendency of individuals to interact with similar ones and, as a consequence, become even more alike. According to this idea, the 
likelihood for an individual to imitate a cultural trait from another individual will depend on 
how many other traits they have already in common. For low values of the initial cultural diversity, the resulting dynamics converges to a
global monocultural state, characterized by agents that share every cultural
trait. In contrast, for high values of initial states multiculturality prevails. This change of macroscopic
behaviour has been characterized as a non-equilibrium phase
transition \cite{castellano2000nonequilibrium,vilone2002ordering,vazquez2007non}.
Subsequent researches studied several issues related to the Axelrod's Model, including the
effects of the network's topology \cite{klemm2003nonequilibrium,guerra2010dynamical}, cultural
drift (modeled as noise) \cite{klemm2003global,klemm2005globalization}, 
local social pressure \cite{kuperman2006cultural}, confidence 
thresholds \cite{de2009effects}, media (represented by an externan field) \cite{gonzalez2005nonequilibrium,gonzalez2007information},
mobility and segregation \cite{gracia2009residential,gracia2011selective}, and dynamic networks \cite{gracia2011coevolutionary}. 
In addition to the Axelrod model, other types of dynamics for vectors of opinions
have been proposed, including binary \cite{deffuant2000mixing,laguna2003vector,biral2015consensus} and 
continuous variables \cite{fortunato2005damage} for the opinions.

Although the Axelrod model can capture some features of societies \cite{valori2012reconciling},
it does not take into account a key characteristic of real-world social dynamics, namely, the
evolution of topics in the social debate. While, for example, in the nineteenth century slavery was discussed and in the first half
of the twentieth century there was a debate on women's suffrage, currently these
themes are not any more at debate. Instead, new issues arise and become the center of the political discourse, as for example, abortion and LGBT rights.

In this work, we consider a model that takes into account the open-ended nature 
of the social debate. This particular aspect of social dynamics has been previously
dealt with in other ABM used to describe the exchange of linguistic
conventions \cite{brigatti2009conventions,crokidakis2015discontinuous}.
In the case of the Axelrod's model, an open set of cultural features 
is easily introduced by substituting the topics which achieve a high degree of consensus.
This is implemented reinitializing with random traits the cultural feature
that achieves a level of agreement greater than a threshold $\phi$.
The parameter $\phi$ can be interpreted as
the resistance of the society, that is,
the minimum level of agreement required to assume consensus on an issue. Our numerical
results show that the emergence of new topics for discussion into the social debate has little effect on cultural
groups once they have been consolidated, but it does have a considerable effect
on cultural overlaps. Along with Monte-Carlo simulations, we have also performed a mean-field analysis. Although the mean-field approach reproduces qualitatively some aspects of the numerical
results, it substantially underestimates the value of the cultural overlap, a fact that highlights 
the influence of the topology and the correlations between the different cultural
features in the Axelrod's dynamics.


\begin{figure*}[ht!]
\begin{center}
\includegraphics[width=6.8in, angle=0]{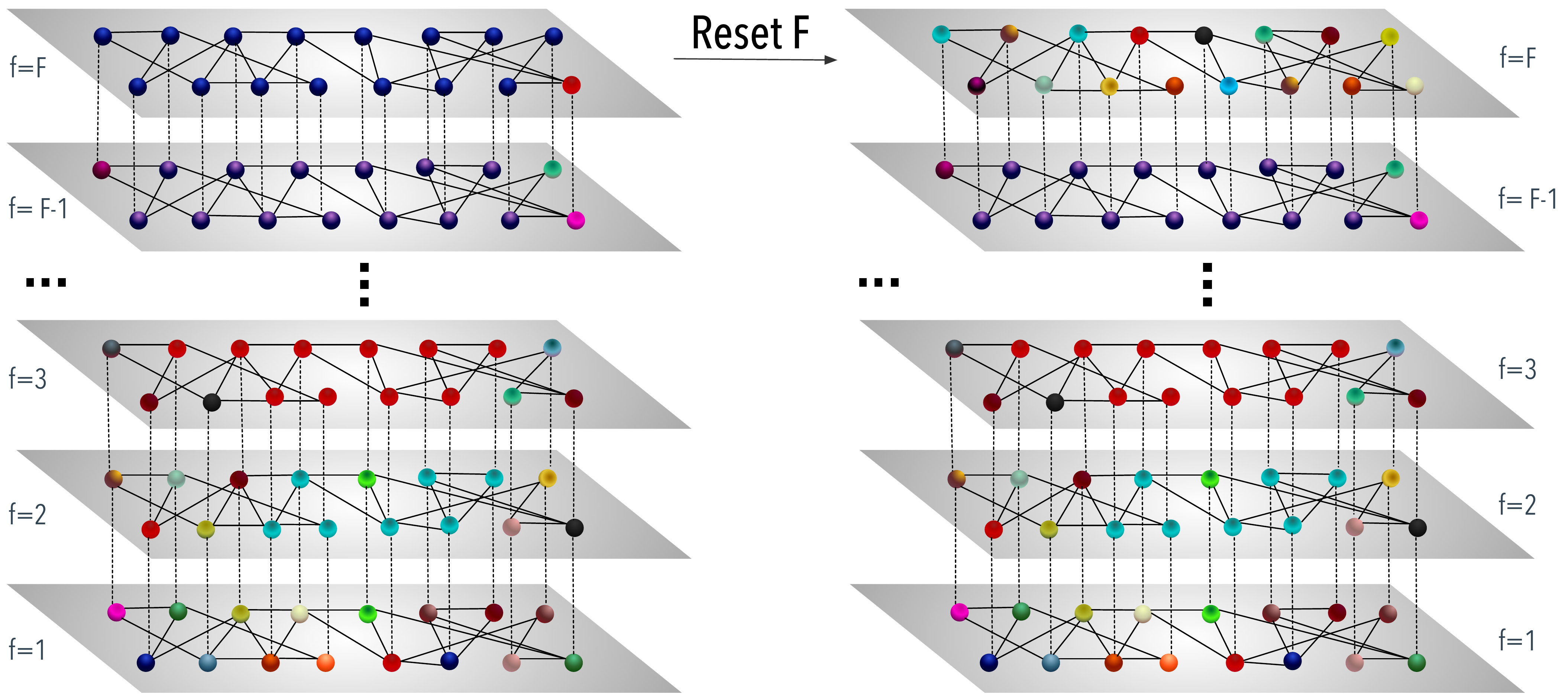}
\end{center}
\caption{(color online) In this illustrative figure, each layer represents a different feature
($f=1,2,\ldots, F$), while nodes represent the agents. Each agent is depicted by the same node in all the layers, and links stand for the contacts between agents. When the fraction of agents sharing the most
abundant trait of a feature reaches the value $\phi$ (left panel, layer $F$), consensus on the topic
is assumed and it is replaced by a new emerging topic through the initialization of
traits in layer $F$ (right panel).}
\label{artWork}
\end{figure*}


\section{Renewal of social debate topics in the Axelrod model}

In the original Axelrod model of cultural dissemination, $N$ cultural
agents occupy the nodes of a network whose links define the social contacts among them. Each
agent $i$ is assigned to a culture modeled as a vector
of $F$ integer variables $\{\sigma_f(i)\}$ ($f=1,...,F$), the {\em cultural features}, that can
assume $q$ values, $\sigma_f= 0,1,...q-1$, the {\em traits} of the feature. 
The features of each agent $i$ are initialized by random assignment of traits from
a uniform distribution. 
The parameter $q$ represents the initial cultural
diversity. At each time step, a random agent $i$ is chosen and allowed
to imitate an unshared feature's trait of a randomly chosen neighbor $j$, with a probability
proportional to their cultural overlap $\omega_{ij}$, which is defined as the fraction
of common cultural features, 
\begin{equation}
\omega_{ij} =
\frac{1}{F}\sum_{f=1}^{F}\delta_{\sigma_f(i),\sigma_f(j)}\;,
\label{overlap}
\end{equation}
where $\delta_{x,y}$ is the Kronecker's delta defined as $\delta_{xy}=1$
if $x=y$ and $\delta_{xy}=0$ otherwise. 

In this work, we consider the incorporation of new topics into the social
debate as consensus is reached in other topics, modeling this situation through the
reinitialization of the features
in which a given level of agreement has been reached. Explicitly, when the proportion
of agents sharing the most
abundant trait of a feature $f$ exceeds a threshold $\phi$ ($0<\phi|\le1$), consensus on the topic
is assumed and the topic is replaced by a new one. 
To this end, see Figure \ref{artWork}, the feature $f$ of each agent
is drawn randomly from a uniform distribution on the integers $\lbrace1,2,\ldots,q\rbrace$. The parameter $\phi$ (here called resistance)
represents the minimum level of agreement required 
to assume consensus on a topic. Note that
for $\phi=1$ the original Axelrod model is recovered.

\section{Results and discussion}

In this section we present the numerical results of our Monte Carlo simulations along with analytical results obtained for a mean field approximation. In order to compare the numerical results with the analytical ones, for the simulations we consider the case
of random regular networks. Random regular networks are random networks of fixed degree $k$, which means that all nodes are equivalent. 

\begin{figure}[ht!]
\begin{center}
\includegraphics[width=2.9in, angle=0]{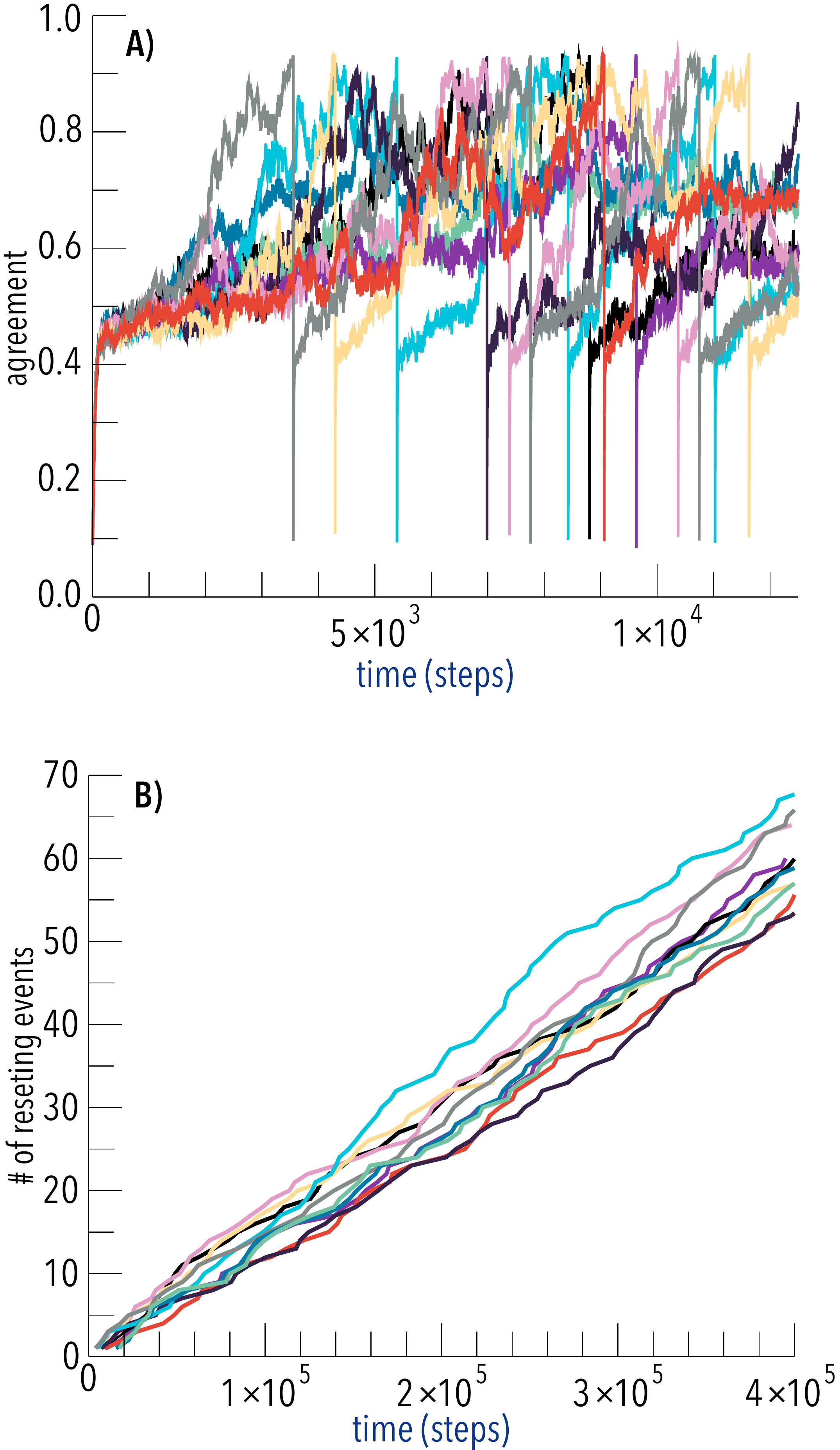}
\end{center}
\caption{(color online) \textbf{a)} Time evolution of the agreement level, \textit{i.e.}, fraction of agents sharing the most
abundant trait, on a given feature. When the level of agreement reaches the value $\phi=0.95$, the feature is
reinitialized by random assignment of traits. \textbf{b)} Cumulative number of resetting
events (feature's reinitialization) as a function of time, for different features. 
The panels show the evolution of a representative realization. Different colors represent different
features. Other values are $q=F=10$, $N=10^3$. See the text for further details.}
\label{DynamicsFigure}
\end{figure}

The system is initialized by a random assignment of agent's cultures, that is, for every node in the network the $F$ features are drawn randomly from a uniform distribution on the set of integers $\lbrace1,2,\ldots,q\rbrace$. 
The process is stopped when the system reaches a stationary state, characterized by quasi-constant values of the observables between resetting events. The results shown below are obtained by averaging over a large number (typically $100$) of random regular networks with $k=6$ and different initial conditions.

To illustrate the dynamics proposed here, Figure \ref{DynamicsFigure} displays the time evolution of a characteristic
realization for $F=10$, $q=10$, and a 
value of $\phi=0.95$. Panel \textbf{a)} shows the evolution of the level of agreement on
each feature, \textit{i.e.}, the fraction of agents sharing the most
abundant trait of a given feature. Different colors represent the different features $f=1,2, \ldots,F$. As shown, 
when the level of agreement on a feature $f$ reaches the value $\phi=0.95$, the feature $f$ is
initialized by assigning at random a new value to the corresponding cultural component in the
cultural vectors of all the agents. This leads to a value close to $1/q$ for the level of agreement on the reseted feature. Panel \textbf{b)} shows the time evolution of the cumulative number of feature's initializations, each line representing a different feature.
As it can be seen, the symmetry of the dynamics, according to which there are not prevalent features,
entails a similar evolution of the cumulative number of resetting events for the different features.

\begin{figure}[ht!]
\begin{center}
\includegraphics[width=2.9in, angle=0]{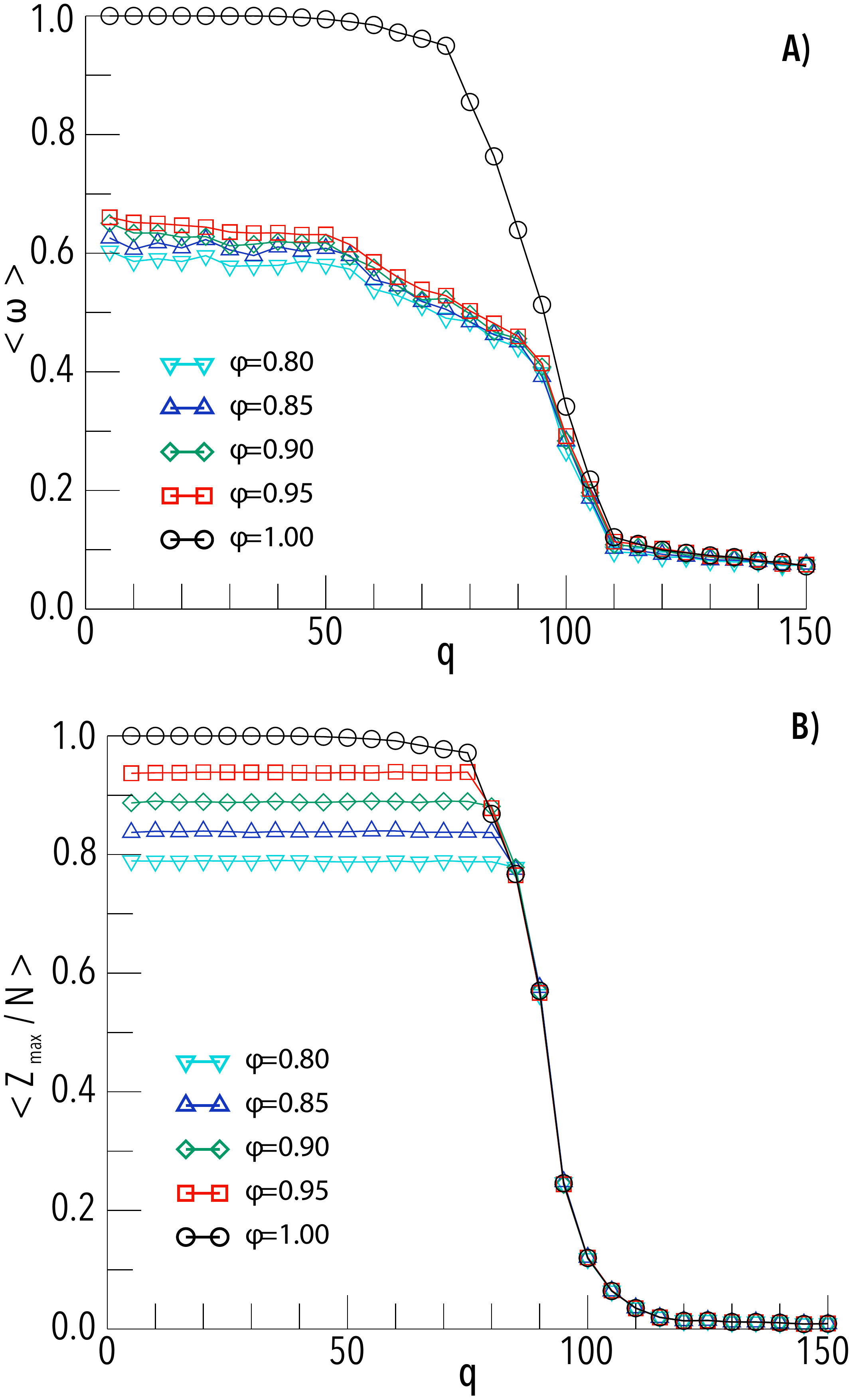}
\end{center}
\caption{(color online) \textbf{a)} Average cultural overlap $\langle\omega\rangle$ in the steady state versus
the number of traits $q$, for different values of the resistance
$\phi$. \textbf{b)} Fraction of agents sharing the most abundant trait $\langle Z_{max}/N \rangle$
in the steady state versus the number of traits $q$, for different values of the resistance
$\phi$. Other values are $F=10$, $N=10^3$. Each point is obtained by averaging over 100 independent realizations. See the text for further details.}
\label{ResetFigure}
\end{figure}

The usual order parameter for the original Axelrod model is $S_{max}/N$, where $S_{max}$ is the average number of agents
of the most abundant culture. Large values (close to unity) of the order parameter represent cultural globalization. In particular, 
in the ordered state ($S_{max}/N=1$) all the agents belong to the same cultural group, that is, they share all the cultural traits. 
Nevertheless, the model here proposed ($\phi<1$) precludes this monocultural state. Actually, due to the open nature of the cultural features it is no longer possible to characterize cultures through unanimous consensus on 
all the topics, being more convenient to impose a less restrictive condition. In this sense, 
the cultural overlap averaged over all the links $\langle \omega \rangle$ constitutes a measure of multiculturalism. The averaged overlap $\langle \omega \rangle$ is defined as:

\begin{equation}
\langle \omega \rangle=\frac{1}{E}\sum_{i}\sum_{j\neq i}A_{ij}\omega_{ij}\;,
\label{averagedOverlap}
\end{equation}
where $E$ is the number of links, $A_{ij}$ is the element $(i,j)$ of the adjacency matrix $-$$A_{ij}=1$ if $i$ and $j$ are linked and 0 otherwise$-$, and $\omega_{ij}$ is the 
cultural overlap of agents $i$ and $j$ defined in (\ref{overlap}). Large values of the average overlap ($\langle \omega \rangle \sim 1$) correspond to a state close to monoculturalism,
while low values ($\langle \omega \rangle \sim 0$) correspond to multiculturalism. In Panel \textbf{a)} of Figure \ref{ResetFigure},
we plot the average cultural overlap $\langle\omega\rangle$ as a function of the initial cultural diversity $q$, for 
different values of the resistance $\phi$. As it can be seen, for resistance values lower than one ($\phi < 1$) all graphs almost collapse in a single curve. Furthermore, by comparing the curves corresponding to the model here proposed ($\phi < 1$) 
with the one corresponding to the original Axelrod model ($\phi =1$), it is observed that the curves for $\phi < 1$ show less overlap for almost any value of $q$, which implies that the emergence of new themes in the social debate has a strong impact on cultural overlap, leading to a decrease of it. 

\begin{figure}[t!]
\begin{center}
\includegraphics[width=2.9in, angle=0]{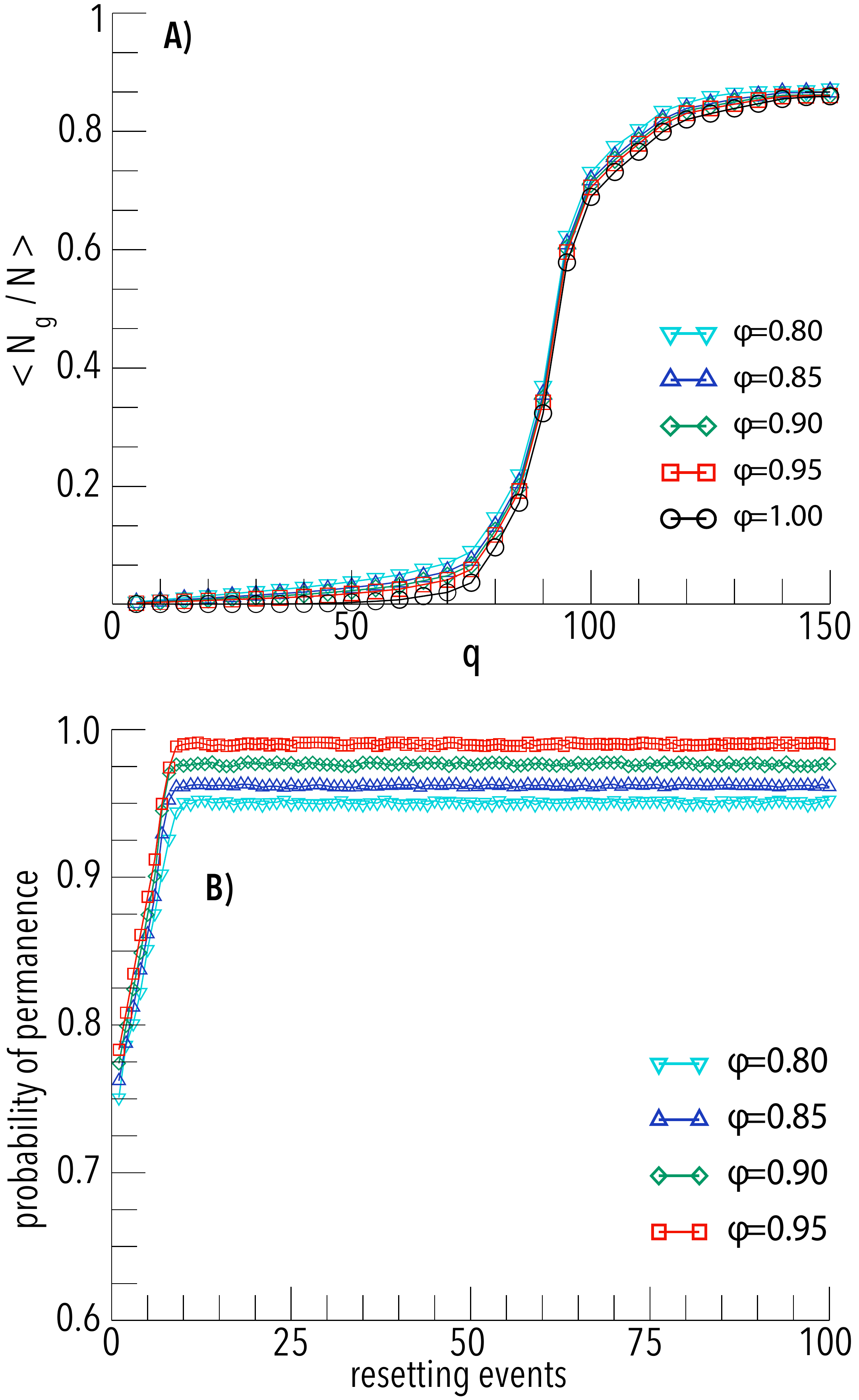}
\end{center}
\caption{(color online) \textbf{a)} Scaled number of different cultural groups $\langle N_g/N \rangle$ versus
the number of traits $q$, for different values of the resistance $\phi$. \textbf{b)} Evolution of the probability for an agent to remain in the same cultural group between two consecutive reset events, for different values of the resistance $\phi$ and for a number of traits $q=80$.
Here, two agents belong to the same cultural group if they share, at least, $F-1$ cultural traits. Other values are $F=10$, $N=10^3$. Each point is obtained by averaging over 100 independent realizations. See the text for further details.}
\label{CulturalGroupsFigure}
\end{figure}

As a complementary observable, we have also computed the fraction of agents sharing the most abundant cultural trait, $ Z_{max} / N$, which measures partial opinion convergence. Here $Z_{max}$ stands for the number of agents
that share the most common trait $q$ of the feature $f$ with the highest level of
agreement. In Panel \textbf{b)} of Figure \ref{ResetFigure},
we plot $Z_{max} /N$ as a function of the initial cultural diversity $q$, for different values of the resistance $\phi$. As it is shown, the only effect of the features' resetting in the partial opinion convergence is limited to low values of $q$ (those corresponding
to the ordered state in the original Axelrod model). On the other hand, this discrepancy for low values of $q$ is
almost the minimum difference that the constraint imposed by the value of $\phi$ allows. This means that
the incorporation of new topics into the social debate has a minimal effect on the convergence
of the rest of the topics in which there is already a high degree of consensus. 

The small effect of the emergence of new debate topics on the rest of the cultural features
of the agents makes it of further interest to study how it affects the dynamics of cultural groups. In the
original Axelrod model, cultural domains are composed of agents that share
all the traits, that is, two agents $i$ and $j$ belong to the same cultural domain if, and only if,
their corresponding cultural vectors are equal $\{\sigma_f(i)\} = \{\sigma_f(j)\}$. 
As exposed above, the model here proposed ($\phi<1$) bans the formation of homogeneous cultural domains and
it is more convenient to impose a less restrictive condition. To this end, we relax the definition of cultural groups by allowing that
two agents belong to the same cultural group if they share, at least, $F-1$ cultural traits. According to the new condition, the resetting of a
feature should not lead to the disintegration of all the groups so defined. The observable $\langle N_g/N \rangle$,
where $N_g$ is the number of cultural groups in the final state, provides a measure of the disorder
of the system \cite{gonzalez2005nonequilibrium}. 

Figure \ref{CulturalGroupsFigure} shows the numerical results for
the cultural groups, as defined above, on a random regular network of size $N=1000$ and
degree $k=6$. In panel \textbf{a)} of
Figure \ref{CulturalGroupsFigure} we show the
average fraction of different cultural groups $\langle N_g / N\rangle$
as a function of the initial cultural diversity $q$, for diferent values of the
resistance $\phi$. As it can be seen, the resistance has little influence on the number of cultural groups in the steady
state, as it is expected that the resetting process does not have a strong effect on the dynamics of the cultural groups. We also show in Panel \textbf{b)} of Figure \ref{CulturalGroupsFigure} the time evolution of the probability $p$ that an agent remain in the same cultural group between two consecutive reset events, for a number of traits ($q=80$) corresponding to the transition between ordered
and disordered phases and for different values of the resistance $\phi$. As it is shown, there is a transient
where the probability to remain in the same group increases with time. This transient corresponds to a number of resets
equal to the number of cultural features $F$. After this transient, the permanence
probability $p$ is constant and close to one ($p>0.95$ for resistance values $\phi\geq 0.8$). This means that, once
the cultural groups have been consolidated, the emergence of new topics in the social
debate does not have a strong effect on them. Furthermore, the higher the
resistance, the greater the probability of permanence and, therefore, the less influence of
the renewal of topics on the structure of the cultural groups.

\subsection*{Mean-field estimation}

The dynamics of the problem can be studied using a mean-field approach \cite{castellano2000nonequilibrium}. Let $P_m(t)$ denote 
the probability that a random link connects two agents that agree on $m$ topics,
that is, with overlap $\omega_{ij}=m/F$.
If the initialization of the traits is random, unbiased and uncorrelated, at time $t_0=0$ we have:

\begin{equation}
P_m(0)=\frac{F!}{m!(F-m)!} \left(\frac{1}{q}\right)^m  \left(\frac{q-1}{q}\right)^{F-m} \;.
\label{initialDistributionMF}
\end{equation}
\\
In the mean-field approach, the time evolution of $P_m(t)$ can be obtained as:

\begin{eqnarray}
\frac{dP_m}{t}&=&\sum_{u=1}^{F-1}\frac{u}{F}P_u \bigg\{  \delta_{m,u+1}-\delta_{m,u}   \nonumber \\
 &+&(\langle k\rangle-1)\sum\nolimits_{n=0}^F(P_nW_{n,m}-P_mW_{m,n}) \bigg\}   \;,
\label{timeEvolutionMF}
\end{eqnarray}
\\

where $\langle k\rangle$ represents the average connectivity and $W_{n,m}$  is the transition
probability for two connected agents sharing $n$ traits to share $m$ traits due to the update of 
a neighbor of one of them. The average overlap $\omega(t)$ can be expressed as:

\begin{equation}
\omega(t)=\sum_{u=1}^F\frac{u}{F}P_u \;.
\label{overlapMF}
\end{equation}
Neglecting correlations among neighboring links, the second-order transition probabilities
are given by:

\begin{equation}
W_{n,m}=\frac{F-n}{F}\;\omega(t)\;\delta_{m,n+1}+\frac{m}{n}\delta_{m,n+1} \;.
\label{transitionProbabilites}
\end{equation}
The first term of equation (\ref{transitionProbabilites}) represents the probability that
two agents with overlap $n/F$ increase their overlap to $(n+1)/F$ due to the update of
a neighbor of one of the two agents, while the second term represents the probability that
those agents decrease their overlap to $(n-1)/F$ due to the update of
a neighbor.

Finally, note that the average overlap $\omega(t)$ can be expressed also as:
\begin{equation}
\omega(t)=\frac{1}{F\sum_{i,j}A_{ij}}\sum_{i,j}A_{ij}\sum_{f=1}^{F}\delta_{\sigma_f(i),\sigma_f(j)}\;,
\end{equation}
where $A_{ij}=1$ if agents $i$ and $j$ are connected and $0$ otherwise. 
Considering the commutativity of summations, we have that
\begin{eqnarray}
\omega(t)=\frac{1}{F\sum_{i,j}A_{ij}}\sum_{i,j}A_{ij}\sum_{f=1}^{F}\delta_{\sigma_f(i),\sigma_f(j)} = \nonumber \\
\frac{1}{F\sum_{i,j}A_{ij}}\sum_{f=1}^{F}\sum_{i,j}A_{ij}\delta_{\sigma_f(i),\sigma_f(j)}\;,
\label{commutativity}
\end{eqnarray}
neglecting the correlations among the different features and provided that the function $\omega(t)$ is monotonically increasing, one can estimate the average overlap in the stationary state $\langle\omega\rangle$ by means of the integral:

\begin{equation}
\langle\omega\rangle\sim\frac{1}{\tau}\int_{0}^{\tau}\omega(t)dt \;,
\label{stationaryAverageOverlapMF}
\end{equation} 
where $\tau$ stands for the value $t=\tau$ such that $\omega(\tau)=\phi$. In the above
estimation we have considered that, after a long
enough transient time, the resetting of the different features has
distributed them homogeneously over time. The time-averaged value of the overlap of
a feature constitutes an estimator of the overlap averaged over all the features. 
This idea relies on the summation exchange shown in equation (\ref{commutativity}),
on disregarding correlations between features and on the symmetry of the model,
that is, the non-prevalence of a feature over the others.

\begin{figure}[t!]
\begin{center}
\includegraphics[width=2.9in, angle=0]{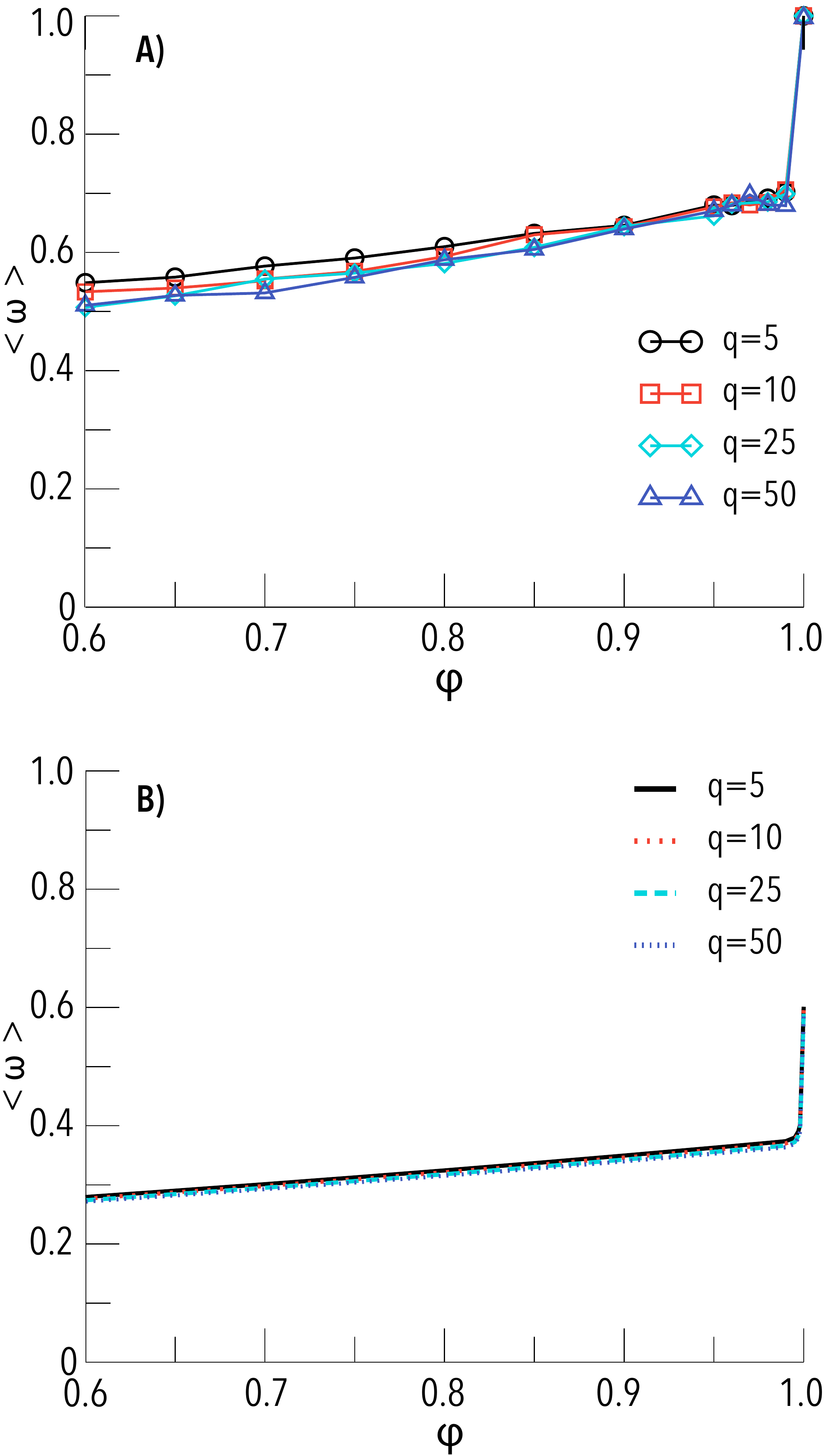}
\end{center}
\caption{(color online)\textbf{a)}  Numerical results of the average overlap $\langle\omega\rangle$ as a function of the the resistance 
$\phi$, for different values of the number of traits $q$. Other values are $F=10$, $N=10^3$, $k=6$. 
\textbf{b)} Mean-field estimation. As shown, mean-field predictions underestimate the cultural overlap. This shows the influence of the network of contacts as well as the correlations between the different cultural features in Axelrod's
dynamics. See the text for further details.}
\label{ResetMFFigure}
\end{figure}

The equations (\ref{timeEvolutionMF}-\ref{stationaryAverageOverlapMF}) can be solved
numerically. Figure \ref{ResetMFFigure} shows the average overlap in the stationary
state $\langle\omega\rangle$ as a function of the the resistance $\phi$, for different values
of the number of cultural traits $q$. The comparison between numerical results (panel \textbf{a}) and theoretical
predictions (panel \textbf{b}) highlights that mean-field approach
underestimates the cultural overlap value. The reasons for this underestimation
in the mean-field approximation rely on the assumptions on which it was based. On one
hand, in the Axelrod dynamics, cultural clusters are associated to topological clusters, 
which are totally absent in the mean-field approximation. On the other hand, the homophily
mechanism enhances correlations among the different features, while the mean-field
approximation neglects those correlations. Note that these two characteristics of the Axelrod 
dynamics, namely the formation of cultural clusters associated with
the contact network and the correlation between the different
cultural features, establish a connection
between the formation of cultures and interpersonal relationships. Notwithstanding the quantitative disagreement between MC simulations and the analytical approximation, the latter does capture the behavior of the model, and thus, provides mechanistic hints about what is going on in the system's dynamics.

\section{Summary and concluding remarks} 

In the Axelrod model for cultural dissemination, we have considered the incorporation of new topics into
the social debate by resetting the features in which the
fraction of agents sharing the most abundant trait exceeds a threshold $\phi$. This parameter $\phi$, that we call resistance,
represents the minimum level of agreement required to assume consensus on a topic.
The introduction of an open-ended set of topics through this
resetting mechanism avoids the frozen monocultural state of the original Axelrod's model. We have
performed extensive numerical simulations which show that, for high enough
values of the initial cultural diversity, the dynamics leads to a multicultural society fragmented
into clusters characterized by a high degree of cultural agreement. 
Remarkably, we show that the renewal of the social discussion topics does not have a considerable effect on
the distribution of consolidated cultural clusters. This preservation of group cohesion is consistent with the idea that individuals take a position on emerging issues of social debate in accordance with the trend of the cultural group they belong to. However, this renewal of discussion topics has a
significant influence on the cultural overlaps, with the result that cultural clusters are less homogeneous than in the case of a closed set of discussion topics.

In addition, we have performed a mean-field analysis based on two assumptions, namely, the presumption that the agents interact with each other in proportion to their average abundance and the disregard of correlations between the different features. Although the mean-field analysis qualitatively reproduces the numerical results, it yields a underestimation of the mean cultural overlap. This underestimation of the mean-field approximation highlights the key role of the local interactions in the Axelrod dynamics, where cultural and topological clusters are closely linked, as well as the imitation-driven correlations among the different cultural features. Altogether, our work introduced opens the path to considering more sophisticated models in which agreement is not frozen once reached and to including higher-order correlations, for instance, by considering updating rules that involve more than one (possibly correlated) features.

\begin{acknowledgments} A. R. H. thanks COSNET Lab at the Institute BIFI for partial support and hospitality during the realization of most of this work. C. G. L. and Y. M. acknowledge support from the Government of Arag\'on, Spain through a grant to the group FENOL, by MINECO and FEDER funds (grant FIS2014-55867-P) and by the European Commission FET-Open Project Ibsen (grant 662725).
\end{acknowledgments}

\bibliography{axelrodReset}{}

\end{document}